# High-Precision Multi-Wave Rectifier Circuit Operating in Low Voltage ± 1.5 Volt Current Mode

Bancha Burapattanasiri

Department of Electronic and Telecommunication Engineering, Engineering Collaborative Research Center
Faculty of Engineering, Kasem Bundit University
Bangkok, Thailand 10250
.

*Abstract*—This article is present high-precision multi-wave rectifier circuit operating in low voltage ± 1.5 Volt current modes by CMOS technology 0.5 µm, receive input and give output in current mode, respond at high frequency period. The structure compound with high-speed current comparator circuit, current mirror circuit, and CMOS inverter circuit. PSpice program used for confirmation the performance of testing. The PSpice program shows operating of circuit is able to working at maximum input current 400 µA$_{p-p}$, maximum frequency responding 200 MHz, high precision and low power losses, and non-precision zero crossing output signal.

## I. INTRODUCTION

The rectifier circuit is very significance in analog signal processing for example, AC voltmeter, detector signal circuit, demodulate circuit etc. [1,2] then, always development all the time for example, full wave rectifier circuit operating in voltage mode [3], full wave rectifier circuit operating in diode and bipolar transistor [4], there are used voltage operating around 0.3 volt for Ge, and 0.6 volt for Si, then it has signal error in crossing zero, and in low input signal case the circuit doesn't work, because of the characteristic of diode and transistor has limited. From the limitation has development to use another active device. In the past of century has a lot of the presentation in rectifier circuit in current mode [5], but the circuit still have complicated, dissipation of current source, it has little precision, responding in low frequency and operating in narrow range current, So, In this article would like to show a new choice of easier rectifier circuit, non complicate in active devices connected, but still have the high performance of working by 0.5 µm CMOS technology, and the compound with high-speed current comparator circuit, current mirror circuit, and CMOS inverter circuit. Thus, it able to responding at high operating precision, working at high frequency, responding at high input current, high precision signal, high speed, low disputation power, and non precision zero-crossing output signal.

Identify applicable sponsor/s here. *(sponsors)*

## II. BASIC PRINCIPLE AND DESIGNATION

### A. Current Mirror Circuit

The current mirror circuit is compound input and output current circuit, it is the current source to amplifier circuit, because of low input resistance but high output resistance that is the characteristic of current source.

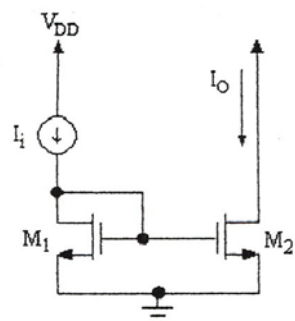

Figure 1. Basic current mirror circuit

From figure 1 $M_1$ pin drain connected to $M_1$ pin gate, which is the same of diode. When current source is stable $I_1$ pass to $M_1$ then voltage drop between pin gate and pin source of $M_1$, thus voltage drop at $M_2$ too, when $M_1$ and $M_2$ has same characteristic then $I_1 = I_{in}$ as equation (1) and (2)

$$I_D = \frac{1}{2} k'_n \frac{W}{L} (V_{gs} - V_t)^2 \qquad (1)$$

$$I_o = I_i \frac{(W/L)_2}{(W/L)_1} \qquad (2)$$

### B. Completely High-Precision Multi-Wave Rectifier Circuit Operating in Low Voltage ± 1.5 Volt Current Mode.

From the principle of current mirror circuit, when designing to rectifier circuit it able to show the structure of completely high-precision multi-wave rectifier circuit operating in low voltage ± 1.5 Volt current mode as figure 2. The structure compound with high-speed current comparator circuit, current mirror circuit, and CMOS inverter circuit. The functional working is when high-speed current comparator circuit received positive and negative input signal, then it is comparisons the current by sender negative signal to $CM_1$, and sender positive signal to $CM_2$, from the CM characteristics as equation (1) and (2), then will be has half-wave positive and negative phase output signal from $CM_1$ and $CM_2$

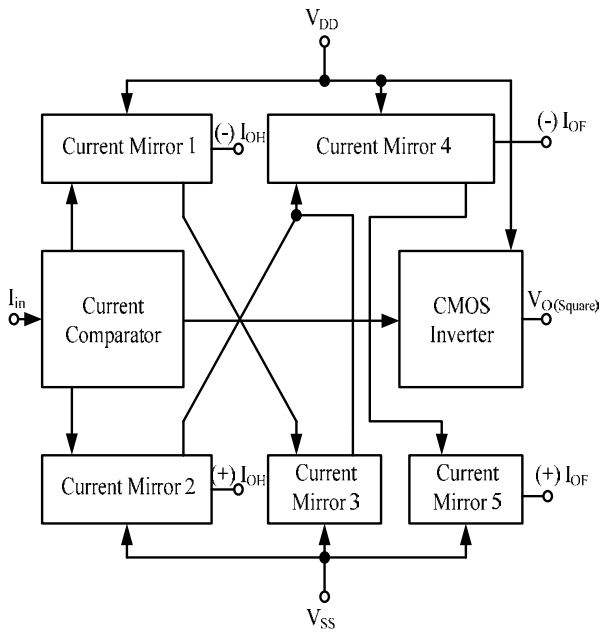

Figure 2. Diagram is show high-precision multi-wave rectifier circuit operating in low voltage $\pm$ 1.5 volt current mode.

While, if send $CM_1$ output signal to $CM_3$ and add it to $CM_2$ output signal, then sending to $CM_4$ input after that it will be full wave negative phase signal. In the similarly things, when $CM_4$ output signal send to $CM_5$ input by the characteristic and the principle of CM as equation (1) and (2), then it will be full-wave negative phase output signal at the same time positive and negative input signal passing to high-speed current comparator circuit, then some part of signal passing to input CMOS inverter circuit for establish square wave signal operation in voltage mode.

*C. Circuit Description*

From figure 2 rectifier circuit, if high-speed current comparator circuit, current mirror circuit, and CMOS inverter circuit put together and sitting MOS transistor operation at saturation region after that it will be completely high-precision multi-wave rectifier circuit operating in low voltage $\pm$ 1.5 volt current mode as figure 3.

$$I_{DM2} = 0 \text{ And } I_{DM1} = I_{in} \text{ When } I_{in} > 0 \quad (3)$$

$$I_{DM1} = 0 \text{ And } I_{DM2} = I_{in} \text{ When } I_{in} < 0 \quad (4)$$

From circuit in figure 3 when sender input current signal as equation (3) after that $M_3$ currenting $M_4$ non currently, output voltage sending back to common source circuit, then $M_1$ currenting and $I_{in} \leq 0$ is equation to $M_2$ pin drain current, and it will mirrored current passing to $M_6$. So, the new signal is half-wave negative phase output signal at $M_6$ pin drain as equation (5).

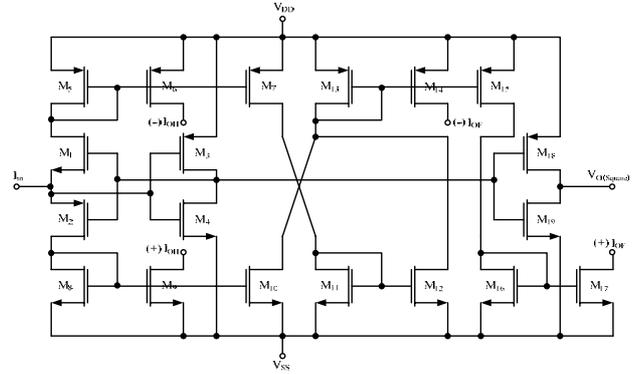

Figure 3. Completely high-precision multi-wave rectifier circuit operating in low voltage $\pm$ 1.5 volt current mode.

In the opposite things, when input current signal as equation (4) will make $M_4$ currenting and $M_3$ non currenting, $V_{ss}$ voltage sender to input common source circuit, then $M_4$ currenting $I_{in} \geq 0$ it equation to $M_2$ pin drain current and it passed to $M_8$ pin drain, reflected current to $M_9$, after that out put signal will be positive phase at $M_9$ pin drain as equation (6).

$$I_{DM5} = I_{DM6} = -I_{OH} \quad (5)$$

$$I_{DM8} = I_{DM9} = +I_{OH} \quad (6)$$

From the characteristic and principle of current mirror circuit, if send $M_7$ pin drain signal (it has equation to $M_6$ pin drain signal) pass to $M_{11}$ pin drain, then reflected to $M_{12}$ for combine with full-wave positive phase current signal at $M_{10}$ pin drain and sent it to $M_{13}$ pin drain then reflected current past to $M_{14}$ after that output signal is full-wave negative phase at $M_{14}$ pin drain as equation (7)

$$-I_{OF} = I_{DM14} = I_{DM10} + I_{DM12} \quad (7)$$

$$+I_{OF} = I_{DM15} = I_{DM17} \quad (8)$$

In the similarly thing, if sending signal at $M_{15}$ pin drain to $M_{16}$ pin drain, then reflected current to $M_{17}$ that output signal is full-wave positive phase at $M_{17}$ pin drain as equation (8), after that output signal is full-wave positive and negative phase input signal send to $M_1$, $M_2$, $M_3$ and $M_4$ then some part of signal at $M_3$ and $M_4$ pin drain has passed to $M_{18}$ and $M_{19}$ for establish square signal at pin drain as equation (9).

$$V_{out(square)} = V_{ss} \text{ then } I_{in} > 0 \text{ and } V_{out(square)} = V_{DD} \text{ then } I_{in} < 0 \quad (9)$$

## III. SIMULATION AND MEASUREMENT RESULT

The PSpice programs used for confirmation the performance of testing , so setting parameter 0.5 µm of MIETEC for PMOS transistor and NMOS $V_{DD}$=1.5 V $V_{SS}$=-1.5 V input current operating at range 0-400 $\mu A_{p-p}$. Figure 4 shows half-wave output signal is sending input signal 400 $\mu A_{p-p}$ at frequency 10, 100, 200 MHz. Figure 5 shows full-wave negative phase output signal is sending input signal 400 $\mu A_{p-p}$ at frequency 10, 100, 200 MHz. Figure 6 shows full-wave positive phase output signal is sending input signal 400 $\mu A_{p-p}$ at frequency 10, 100, 200 MHz. Figure 7 shows square-wave output signal is sending input signal 400 $\mu A_{p-p}$ at frequency 10, 100, 200 MHz. Figure 8 shows the characteristic DC current at input signal 400 $\mu A_{p-p}$ and temperature $25^O, 50^O, 75^O$ and $100^O$

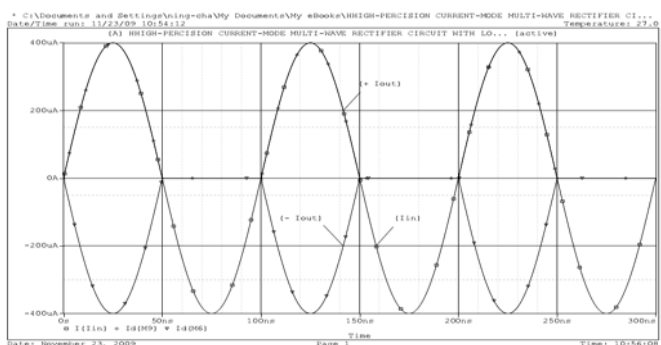

(a)

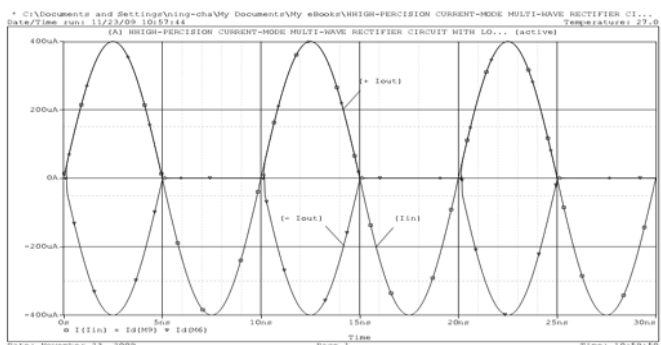

(b)

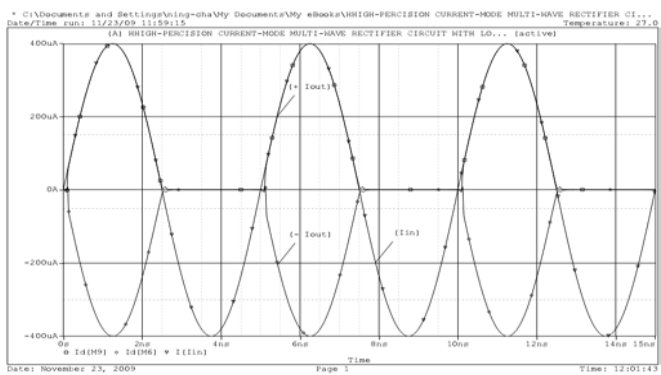

(c)

Figure 4. Half-wave output signal at input current 400 $\mu A_{p-p}$ at (a) frequency = 10 MHz, (b) frequency = 100 MHz and (c) frequency = 200 MHz

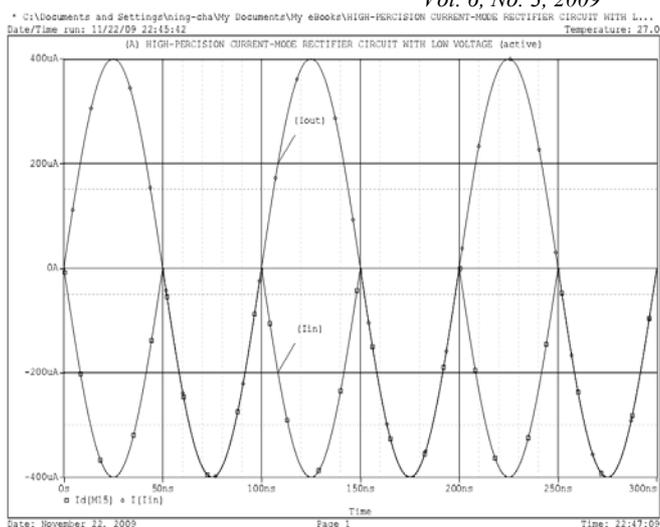

(a)

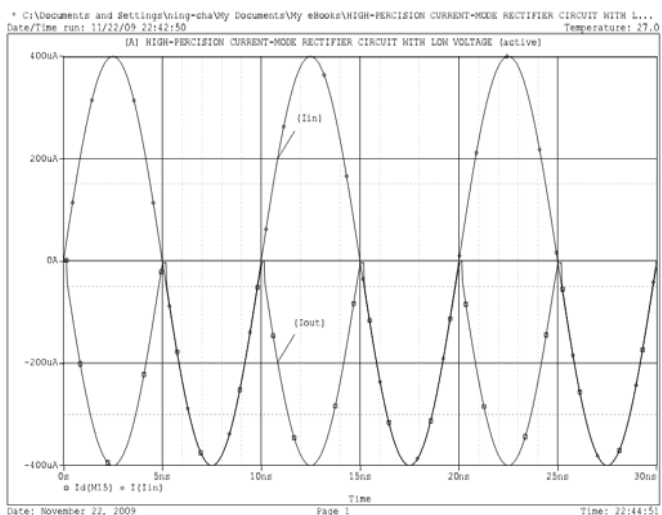

(b)

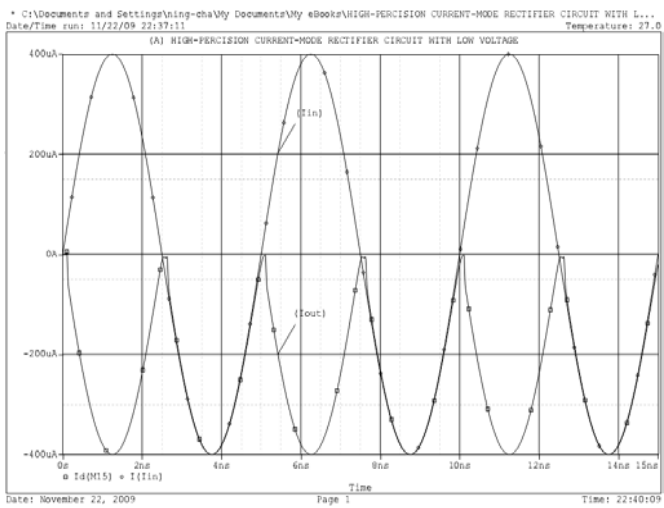

(c)

Figure 5. Full-wave negative phase output signal at input current 400 $\mu A_{p-p}$ at (a) frequency = 10 MHz, (b) frequency = 100 MHz and (c) frequency = 200 MHz

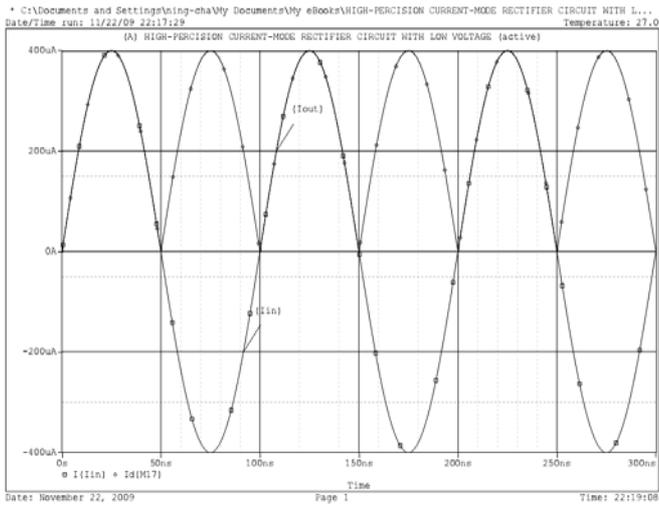
(a)

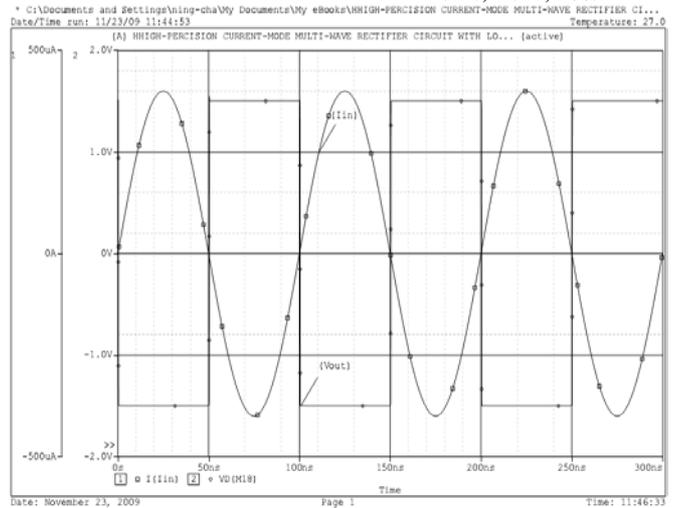
(a)

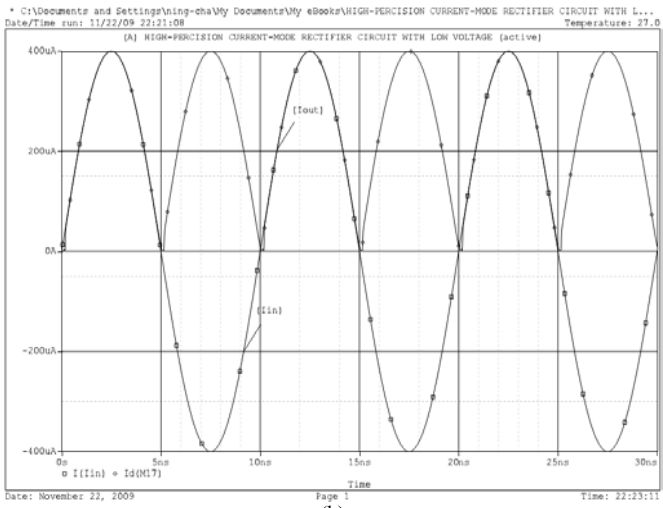
(b)

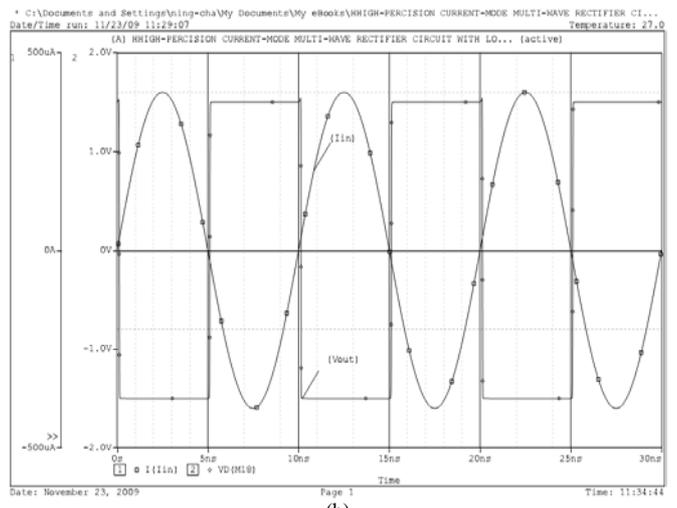
(b)

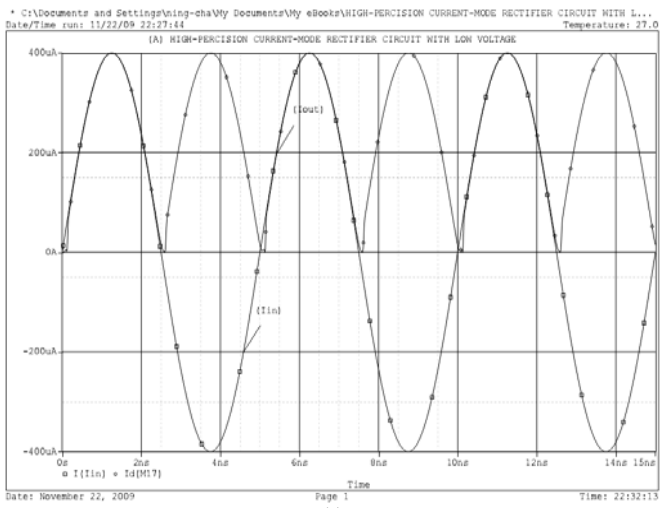
(c)

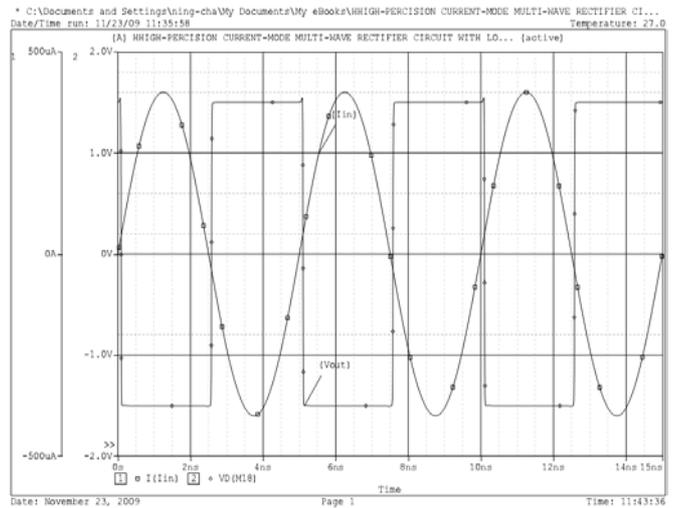
(c)

Figure 6. Full-wave positive phase output signal at input current 400 μA$_{p\text{-}p}$ at (a) frequency = 10 MHz, (b) frequency = 100 MHz and (c) frequency = 200 MHz

Figure 7. Square-wave output signal at input current 400 μA$_{p\text{-}p}$ at (a) frequency = 10 MHz, (b) frequency = 100 MHz and (c) frequency = 200 MHz

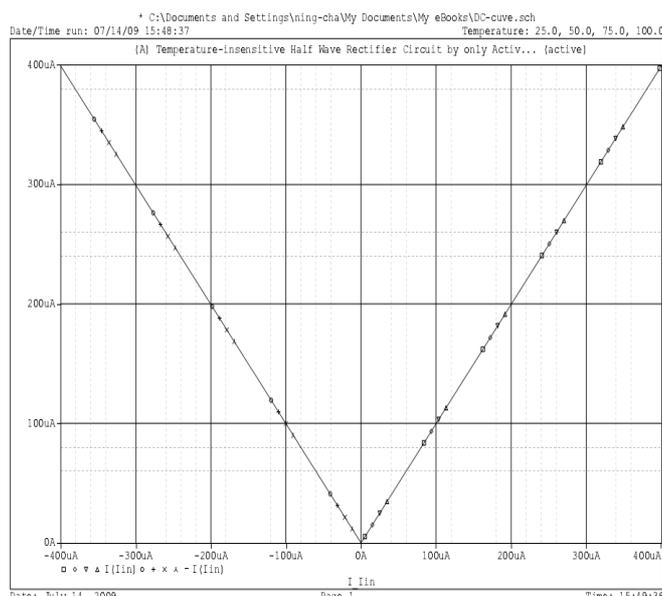

Figure 8. Output characteristic DC current at 400 $\mu A_{p\text{-}p}$ input current, temperature $25^O, 50^O, 75^O$ and $100^O$

## IV. CONCLUSION

The circuit has designed by high-speed current comparator circuit, current mirror circuit and CMOS Inverter circuit. Setting transistor operating at saturation region, so the circuit is able to high precision, but low voltage at $\pm$ 1.5 Volt. The simulation result is able to confirm the performance of working at maximum frequency 200 MHz, maximum output 400 $\mu A_{p\text{-}p}$, high precision, dissipation loss power and non precision zero-crossing output signal, so it suitable to apply in analog signal processing.

## APPENDIX

The parameters used in simulation are 0.5 $\mu m$ CMOS Model obtained through MIETEC [10] as listed in Table I. For aspect ratio (W/L) of MOS transistors used are as follows: 1.5 $\mu m$/0.15 $\mu m$ for all NMOS transistors; 1.5 $\mu m$/0.15 $\mu m$ for all PMOS transistors.

TABLE I.   CMOS MODEL USED IN THE SIMULATION

```
-------------------------------------------------------------------------------
.MODEL CMOSN NMOS LEVEL = 3 TOX = 1.4E-8 NSUB = 1E17
GAMMA = 0.5483559 PHI = 0.7 VTO = 0.7640855 DELTA = 3.0541177
UO = 662.6984452 ETA = 3.162045E-6 THETA = 0.1013999
KP = 1.259355E-4 VMAX = 1.442228E5 KAPPA = 0.3 RSH = 7.513418E-3
NFS = 1E12 TPG = 1 XJ = 3E-7 LD = 1E-13 WD = 2.334779E-7
CGDO = 2.15E-10 CGSO = 2.15E-10 CGBO = 1E-10 CJ = 4.258447E-4
PB = 0.9140376 MJ = 0.435903 CJSW = 3.147465E-10 MJSW = 0.1977689

.MODEL CMOSP PMOS LEVEL = 3 TOX = 1.4E-8 NSUB = 1E17
GAMMA = 0.6243261 PHI = 0.7 VTO = -0.9444911 DELTA = 0.1118368
UO = 250 ETA = 0 THETA = 0.1633973 KP = 3.924644E-5 VMAX = 1E6
KAPPA = 30.1015109 RSH = 33.9672594 NFS = 1E12 TPG = -1 XJ = 2E-7
LD = 5E-13 WD = 4.11531E-7 CGDO = 2.34E-10 CGSO = 2.34E-10
CGBO = 1E-10 CJ = 7.285722E-4 PB = 0.96443 MJ = 0.5
CJSW = 2.955161E-10 MJSW = 0.3184873
-------------------------------------------------------------------------------
```

ACKNOWLEDGMENT

The researchers, we are thank you very much to our parents, who has supporting everything to us. Thankfully to all of professor for knowledge and a consultant, thank you to Miss Suphansa Kansa-Ard for her time and supporting to this research. The last one we couldn't forget that is Kasem Bundit University, Engineering Faculty for supporting and give opportunity to our to development in knowledge and research, so we are special thanks for everything.

AUTHORS PROFILE

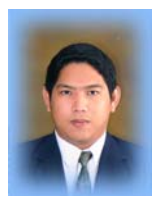

Mr.Bancha Burapattanasiri received the bleacher degree in electronic engineering from Kasem Bundit University in 2002 and master degree in Telecommunication Engineering, from King Mongkut's Institute of Technology Ladkrabang in 2008. He is a lecture of Electronic and Telecommunication Engineering, Faculty of Engineering, Kasem Bundit University, Bangkok, Thailand. His research interests analog circuit design, low voltage, high frequency and high-speed CMOS technology.